\documentclass[12pt]{article}
\usepackage{graphics,cite}
\begin{document}

\begin{titlepage}

\begin{flushright}
\end{flushright}

\vspace{1.cm}

\begin{center}
{\Large \bf\boldmath
Angular-ordered parton showers\\
\vspace{2.5mm}
with medium-modified splitting functions}
\end{center}

\vspace{2mm}

\begin{center}
{\large \bf N\'estor Armesto$^1$, Gennaro Corcella$^{2,3,4}$,\\ \vspace{1.5mm}
Leticia Cunqueiro$^5$ and Carlos A.~Salgado$^1$}\\

\vspace{5mm}

{$^1${\sl Departamento de F\'isica de Part\'iculas 
and IGFAE,}\\ 
{\sl Universidade de Santiago de Compostela, E-15706 Santiago
de Compostela, Galicia-Spain}}
\vspace{3mm}

{$^2${\sl Museo Storico della Fisica e Centro Studi e Ricerche 
E.~Fermi,\\
Piazza del Viminale 1, I-00184 Roma, Italy}}
\vspace{3mm}

{$^3${\sl Scuola Normale Superiore,
Piazza dei Cavalieri 7, I-56126, Pisa, Italy}}
\vspace{3mm}

{$^4${\sl INFN, Sezione di Pisa, Largo Fibonacci 3,
I-56127, Pisa, Italy}}
\vspace{3mm}

{$^5${\sl INFN, Laboratori Nazionali di Frascati,\\
Via E.~Fermi 40, I-00044 Frascati (Roma), Italy}}

\end{center}

\par \vspace{2mm}
\begin{center}
{\large \bf Abstract}
\end{center}
\begin{quote}
  \pretolerance 10000
Modified Altarelli--Parisi 
splitting functions were recently proposed to model 
multi-parton radiation in a dense medium and describe jet quenching,
one of the most striking features of heavy-ion collisions.
We implement medium-modified splitting functions in the
HERWIG parton shower algorithm, which satisfies
the angular ordering prescription, and
present a few parton-level results, such as transverse momentum,
angle and energy-fraction distributions,
which exhibit remarkable medium-induced effects.
We also comment on the comparison with respect to the results yielded by
other implementations of medium-modified 
splitting functions in the framework of virtuality-ordered
parton cascades.
\end{quote}
\end{titlepage}

\section{Introduction}
Measurements performed at the Relativistic Heavy Ion Collider
(RHIC) have emphasized the jet-quenching phenomenon, namely 
the suppression of particle production at large transverse momentum ($p_T$)
with respect
to the naive vacuum expectations, as well as other related phenomena,
such as the disappearance or distortion of the particles directed in opposite
azimuth to a triggered high-transverse-momentum one \cite{brook}.
A typical explanation of jet quenching consists in assuming a higher 
radiative energy loss in a dense medium, which, with respect to the
vacuum case, allows partons potentially 
produced at large $p_T$ to further  emit,
in such a way to decrease the high-$p_T$ multiplicity and enhance the
low-$p_T$ one.   
A lot of work has been undertaken in order to
theoretically describe such effects \cite{quench}.
Calculations carried out so far
are suitable to predict inclusive quantities, but they are not
adequate to describe exclusive final states, which is compelling for the
experimental studies. 
Moreover, they deal with multi-parton emissions only by means of
simple assumptions \cite{dok,salg}.

As happens when studying hadron collisions in the vacuum, Monte
Carlo event generators are  
the best possible tool for the sake of performing 
experimental analyses. In fact, Monte Carlo codes, such as the multi-purpose
HERWIG \cite{herwig} and PYTHIA \cite{pythia} 
generators, contain fairly large
libraries of hard-scattering processes, as well as initial- and final-state 
parton showers, models for hadronization and underlying event.
Moreover, these programs provide one with exclusive final states:
the user can set acceptance cuts on final-state particles
and interface the Monte Carlo output with detector simulation programs.
For such reasons, having Monte Carlo programs capable of simulating
nucleus-nucleus interactions will be of great interest for the 
heavy-ion community working at RHIC \cite{brook} and ultimately at 
the LHC (see Refs.~\cite{alice,cms,atlas} from ALICE, CMS and ATLAS
experiments, respectively).

Although an algorithm for multiple radiation in a dense medium
should require several a priori assumptions, it 
was recently proposed \cite{polosa,mmff} that a simple prescription
to implement medium-induced effects in 
parton shower simulations consists in
adding to the Altarelli--Parisi splitting function a term depending 
on the parameters which characterize the medium,
on the virtuality and on the energy of the branching parton. 
Under reasonable hypotheses on the medium properties,  
it was shown in \cite{polosa} that, even allowing only
one medium-modified splitting, it was possible to obtain azimuthal 
distributions qualitatively similar to the RHIC observations.
Modified splitting functions have been recently implemented in the framework
of the PYTHIA event generator \cite{pythia}. This implementation
is known under the name of Q-PYTHIA: 
the inclusion of modified splitting functions
is detailed in \cite{acs}, whereas the Q-PYTHIA fortran code 
can be downloaded from \cite{qpyt}.
Although a thorough comparison with RHIC data has not yet been performed,
the results of Ref.~\cite{acs} look in qualitative agreement with the
expected features of heavy-ion collisions, namely 
suppression of particle production at large $p_T$, broader angular
distributions and enhancement of intra-jet multiplicities.

In this paper we extend the method discussed in \cite{polosa,mmff,acs}
and apply it to the HERWIG generator \cite{herwig}, the other 
general-purpose Monte Carlo program pretty much used by the experimental 
collaborations. 
In fact, having at least two codes implementing medium-modified splitting
functions is mandatory for the sake of comparison and estimating the
Monte Carlo uncertainty on a prediction.
Also, 
it is well known that the algorithms implemented in HERWIG and PYTHIA are
indeed quite different, in both parton showers and hadronization, which
makes compelling including medium effects even in HERWIG
and using it to analyse heavy-ion collisions.

In fact, HERWIG showers satisfy angular ordering of
multiple soft radiation, which correctly accounts for colour
coherence and allows one to probabilistically implement
multiple soft emissions \cite{marweb}.
On the contrary, 
PYTHIA  cascades are traditionally ordered 
according to the virtuality of the splitting parton, with an option to
reject non-angular-ordered emissions. This was the scenario in which 
the authors of Ref.~\cite{acs} implemented modified splitting functions.
The PYTHIA evolution variable is not,
however, entirely
equivalent to angular ordering in the soft limit:
although in several cases the actual ordering variable does not make
really big changes, when comparing with experimental observables sensitive to
colour coherence, as done in Ref.~\cite{cdf}, HERWIG agrees with the data  
better than PYTHIA.
Transverse-momentum ordering, included in the latest PYTHIA
version \cite{pythiakt}, 
should yield a better description of angular ordering. 
Nevertheless,
as discussed in \cite{bcd}, discrepancies with respect to HERWIG are still
present when considering, e.g., the so-called non-global observables 
\cite{ng}, sensitive to radiation in a limited part of phase space,
such as the transverse energy flow in a rapidity gap.
Even for the purpose of 
hadronization, the two programs implement very different
models, namely the cluster model \cite{cluster} (HERWIG), based on
colour preconfinement, and the string model \cite{string}
(PYTHIA), both depending on a few parameters to
be fitted to experimental data. It is only after turning hadronization on
and fitting such models to the same data set (see, e.g. the study
\cite{bfrag} for heavy-quark fragmentation) that the
comparison between HERWIG and PYTHIA can be consistently made.
At parton-level, due to the above mentioned differences in the
treatment of parton showers, discrepancies between HERWIG and PYTHIA 
must instead be expected, in both vacuum and medium-modified cascades.

More generally, it is worthwhile pointing out that the choice of the ordering
variable for multi-parton radiation in a medium is currently an open
issue (see the discussion in \cite{mmff}).
For single-gluon emission, it is known that finite medium-size effects
affect the radiation pattern by producing a suppression of
small-angle emission, equivalent to the Landau-Pomeranchuk-Migdal
effect \cite{lpm}. Similar mechanisms may, for example, modify the
suppression associated with colour coherence, which leads to angular 
ordering. Implementing medium-modified splitting functions
in angular-ordered parton showers, as we shall do in this paper, 
or in virtuality-ordered cascades, as done in \cite{acs}, 
will allow one to study how the jet structure in a medium
depends on the adoption of a given evolution variable.
Other approaches have also been pursued, following
analytic or Monte Carlo methods \cite{othermc}; the experience 
gained by applying all these different calculations and algorithms 
will be essential for a correct characterization
of the underlying dynamics when experimental data from
heavy-ion collisions will become available.

Hereafter, we shall employ the
latest fortran version of HERWIG, with a few routines modified to
include medium effects. The documentation of this code is 
currently in progress \cite{qher}: in the near future,  
the heavy-ion community will be able to compare the
results of Q-PYTHIA with the ones of the medium-modified Q-HERWIG code.
Furthermore, our work can
be straightforwardly implemented even in
the object-oriented C++ versions of
HERWIG \cite{hwpp} and PYTHIA \cite{pythiapp}, containing the basic 
physics of the fortran ones, plus a number of
remarkable improvements. 

The plan of the present paper is the following.
In section 2 we shall shortly review the basics of the HERWIG 
showering algorithm in the vacuum and how it differs with respect to
the PYTHIA one.
In section 3 we will discuss the  implementation of 
the modified Altarelli--Parisi splitting functions and present
results for the HERWIG Sudakov form factor.
In section
4 we shall present a few parton-level results, showing the role played by 
medium effects. In section 5 we will summarize the
main points of our investigation and make remarks on possible extensions 
of the analysis here presented.

\section{The HERWIG parton shower algorithm}
In this section we discuss the HERWIG algorithm
in the vacuum, to which we shall apply later on the modifications
due to a dense medium.
Monte Carlo algorithms rely on the factorization of the branching
probability for soft or collinear parton radiation.
Referring to final-state radiation, the one mostly affected by 
medium effects, 
the probability of emission of a soft/collinear parton reads \cite{marweb}:
\begin{equation}
  \label{elem}
  d{\cal P}={{\alpha_S}\over{2\pi}}{{dQ^2}\over{Q^2}}\ 
  P(z)\  dz\ 
 {{ \Delta_S(Q^2_{\mathrm{max}},Q_0^2)}\over {\Delta_S(Q^2,Q^2_0)}}.
\end{equation}
In (\ref{elem}), $P(z)$ is the Altarelli--Parisi splitting
function, 
$z$ the energy fraction of the radiated parton with respect to the
emitter and $Q^2$ is the ordering variable of the shower, whose
maximum and minimum values are $Q^2_{\mathrm{max}}$ and $Q^2_0$, respectively.
$Q^2_{\mathrm{max}}$ is set by the hard-scattering process, whereas 
$Q^2_0$ is the user-defined scale at which the perturbative cascade ends and 
cluster hadronization starts \cite{cluster}.

In HERWIG \cite{herwig} the evolution variable is
$Q^2=E^2\zeta$, where $E$
is the splitting-parton energy 
and $\zeta=(p_1\cdot p_2)/(E_1E_2)$, 
$p_1 (E_1)$ and $p_2(E_2)$ 
being the momenta (energies) of the radiated partons.
For massless partons, $Q^2$ turns out to be an energy-weighted angle,
i.e. $Q^2\simeq E^2(1-\cos\theta)$,
where $\theta$ is the emission
angle. For soft radiation, the HERWIG evolution
variable is thus equivalent to angular ordering \cite{marweb},
valid in soft approximation, for azimuthally-averaged quantities, in
the large-$N_C$ limit.
The scale of the strong coupling constant 
$\alpha_S$ is the transverse momentum of the radiated parton
with respect to the emitter. In this way, one includes in the
algorithm a class of subleading soft/collinear logarithms, thus improving
the accuracy of the shower even beyond the leading-logarithmic approximation
\cite{cmw}.

In (\ref{elem}) $\Delta_S(Q^2_1,Q^2_2)$ is the Sudakov form
factor, expressing the probability of evolution from $Q^2_1$ to
$Q^2_2$ 
with no resolvable emission;
in diagrammatic terms, the Sudakov form factor
sums up virtual and unresolved real emissions to all orders.
The ratio of form factors in Eq.~(\ref{elem}) 
represents the probability that the emission at $(z,Q^2)$
is the first, i.e. that there is no radiation during the evolution between 
$Q^2_{\mathrm{max}}$ and $Q^2$. 
It is given by the following expression:
\begin{equation}\label{sudakov}
\Delta_S(Q^2_{\mathrm{max}},Q^2)=\exp\left\{-\int_{Q^2}^{Q^2_\mathrm{max}}
{\frac{dk^2}{k^2}}
\int_{z_{\mathrm{min}}}^{z_{\mathrm{max}}}{dz\ \frac{\alpha_S(z,k^2)}{2\pi}\ 
P(z)}\right\}.
\end{equation}
In Eq.~(\ref{sudakov}), 
the limits of the $z$-integration are
given in HERWIG by \footnote{As discussed in 
\cite{acs}, the $z$-limits in PYTHIA, for a branching with virtuality $p^2$,
are given by $z_{\mathrm{min}}=p^2_0/p^2$ and 
$z_{\mathrm{max}}=1-z_{\mathrm{min}}$, $p_0^2$ being the infrared cutoff.
This yields on
average a larger $z$-evolution range for PYTHIA.}
$z_{\mathrm{min}}=Q_0/Q$ and $z_{\mathrm{max}}=1-z_{\mathrm{min}}$. 

The showering algorithm (\ref{elem}) is frame-dependent, however one can
prove \cite{marweb} 
that, if $i$ and $j$ are colour-connected partons, the upper values
of the evolution variable satisfy the relation $Q_{i,\mathrm{max}}
Q_{j,\mathrm{max}}=p_i\cdot p_j$, which is Lorentz-invariant.
Hence, symmetric choices are made and 
$Q_{i,\mathrm{max}}=Q_{j,\mathrm{max}}=\sqrt{p_i\cdot p_j}$.
\footnote{Such a choice implies, e.g., that for
$e^+e^-\to q\bar q$ annihilation at energy $\sqrt{s}$, the initial
value of the ordering variable for $q$ and $\bar q$ is 
$Q_{\mathrm{max}}=\sqrt{s/2}$.}
Furthermore, the energy of the parton which initiates the shower is
fixed to $E_{\mathrm{max}}=Q_{\mathrm{max}}$: in this way, one identifies the 
HERWIG showering frame. Setting $Q^2<Q^2_{\mathrm{max}}$ for the first
emission yields $\zeta<1$ ($\theta<\pi/2$ in the massless approximatiom)
in the showering frame.


Although in the following we shall just deal with soft or collinear
parton emissions, we point out that 
radiation of hard and large-angle partons can be implemented
in HERWIG by applying matrix-element corrections.
The region
$\zeta>1$, corresponding to hard and large-angle emissions, is the so-called
`dead zone' of the shower, where the exact amplitude is applied.
Moreover, the `hardest-so-far' emission in HERWIG showers is simulated 
according to the exact matrix element, thus allowing a smooth transition
through the boundary of the dead zone \cite{meh}.


Before discussing the implementation of medium effects, we wish to make 
a few comments on the difference between HERWIG and PYTHIA showering
algorithms. 
As discussed in the introduction, PYTHIA includes colour coherence only 
partially: in \cite{acs}, medium modifications were 
included in virtuality-ordered showers, with an option to reject
emission which do not fulfil angular ordering. 
As for matrix-element corrections, PYTHIA has no dead zones: it
uses the soft/collinear
approximation throughout all physical phase space and simulates only the
first branching via the exact amplitude \cite{mep}.
Hadronization is finally implemented in PYTHIA
following the string model \cite{string}.
Since parton showers and possible matrix-element matching are implemented in a
pretty different way, if one studies parton-level quantities, as will be
done hereafter, HERWIG results should not necessarily agree with the
PYTHIA ones presented in \cite{acs}, even though one models 
medium effects in the same way.

Of course, we are aware that a comparison of the results yielded by HERWIG and 
PYTHIA is compulsory. However, it can be meaningfully done only at 
hadron-level and after both generators are tuned to the same data.
In this work, we shall restrict ourselves to comment on possible
discrepancies at parton level and leave to future work a more
detailed comparison, which will demand the redoing of 
the previous extensive studies in
hadron-hadron collisions. 

\section{Medium-modified splitting functions}

In this section we discuss the main issues 
concerning the implementation of
medium effects in parton shower algorithms.
We follow the method presented in \cite{polosa,mmff} and add to 
the Altarelli--Parisi splitting function in the vacuum $P(z)$ a
medium-dependent term $\Delta P(z,p^2,E,\hat q,L)$:
\begin{equation}\label{dp}
P(z)\to P(z)+\Delta P(z,p^2,E,\hat q,L).
\end{equation}
In (\ref{dp}), $p^2$ is the branching-parton virtuality, $E$ its energy,
$L$ the medium length, $\hat q$ the transport coefficient, defined as
the average transverse momentum transferred from the medium to the
parton per unity of free path \cite{dok}.
Hereafter, we shall often make use of the quantities 
$\hat qL$, the so-called accumulated transverse momentum, and 
of the frequency $\omega_c=\hat qL^2/2$. 

The transformation (\ref{dp}) will be applied to the branching algorithm
(\ref{elem}) and, in particular, 
to the integrand function of the Sudakov form factor (\ref{sudakov}),
taking care that
HERWIG evolution variable is not the virtuality $p^2$, but the
energy-weighted angle $Q^2$ defined above 
\footnote{For soft/collinear radiation, it is straightforward to
show that the relation $p^2\simeq 2z(1-z)Q^2$ holds.}. 
As in \cite{acs}, a crucial hypothesis on which our work is based
is that, even in a dense medium, 
the factorization (\ref{elem}) between branching and
no-branching probability still holds and that the showering
evolution variable is the same as in the vacuum. 
In HERWIG, this means that we shall assume angular ordering and colour
coherence even for parton cascades in a medium. In other words, 
the colour flow will not be modified with respect
to its vacuum pattern.

Following \cite{mmff}, we shall modify the splitting functions
for the branchings $q\to qg$ and $g\to gg$, while we shall assume
that the splitting $g\to q\bar q$ gets negligible medium
modifications.\footnote{In any case, the splitting
$g\to q \bar q$ is quite uncommon in the shower, since the
splitting function $P_{qg}(z)=T_R[z^2 +(1-z)^2]$ is not soft-enhanced.
It is only after turning cluster hadronization on 
\cite{cluster} that non-perturbative
$g\to q\bar q$ splittings are forced and $P_{qg}(z)$ plays
a role.}
$\Delta P(z,p^2,E,\hat q,L)$
can be expressed in terms of the medium-induced parton radiation spectrum
$I^{\mathrm{med}}(z,p^2)$, computed in the so-called BDMPS
approximation \cite{bdmps,zak,wiede}:
\begin{equation}\label{med}
\Delta P(z,p^2,E,\hat q,L)\simeq \frac{2\pi p^2}{\alpha_S}
\frac{d^2I^{\mathrm{med}}}{dzdp^2}.
\end{equation}
As discussed in \cite{acs}, Eq.~(\ref{med}) gives the medium corrections
to the soft-divergent part of the Altarelli--Parisi splitting
functions; the finite terms are assumed to be 
vacuum-like\footnote{That means that, 
e.g., for a $q\to qg$ branching, 
where the Altarelli--Parisi splitting function 
is given by $P_{gq}(z)=C_F[1+(1-z)^2]/z$,
medium-induced modifications affect the overall $\sim 1/z$ factor,
while the finite-$z$ corrections remain unchanged.}.

$\Delta P(z,p^2,E,\hat q,L)$ can be expressed in terms of 
the energy $E$ of the parent parton and
of the dimensionless variables $E/\omega_c$ and $\kappa^2=k_T^2/(\hat qL)$,
where
$k_T$ is the transverse momentum of the radiated soft/collinear parton
with respect to the splitting one.
For multiple radiation, 
$E$ will always be the energy of the splitting quark/gluon.
As for the medium length, if 
$L_0$ is the length for the first splitting, for the following emissions
we have to take into account that a radiated parton, with energy fraction
$z$, travels for a distance $2 z E/k_T^2$, the so-called parton formation
length, before splitting again.
Therefore, the effective medium length for subsequent branchings will be:
\begin{equation}\label{ell}
L=L_0-\frac{2zE}{k_T^2}.
\end{equation}
We note that in Eq.~(\ref{ell}) $L$ is not positive definite and,
especially for small values of $L_0$ and very soft/collinear splittings, 
it may well become negative after few emissions. Whenever this is the case,
we shall assume that the shower continues in a vacuum-like fashion.

In Ref.~\cite{mmff} the modified Altarelli--Parisi splitting functions
were implemented in a Sudakov form factor computed in the same way as
PYTHIA does.
It was indeed found that medium effects have a remarkable impact and,
for any given values of $\hat q$ and $L$, the Sudakov form factor is
suppressed with respect to the vacuum one in the full $p^2$-range,
for both quarks and gluons.
As a decreasing of the form factor corresponds to an enhancement of
the branching probability, a higher parton multiplicity in the 
medium-modified shower 
must be expected.
This was indeed found in \cite{acs} using Q-PYTHIA: although for a
consistent comparison with data one must turn hadronization on, such 
an observation is in qualitative 
agreement with the expectations from radiative energy loss \cite{brook}.

\begin{figure}[b]
\centerline{\resizebox{0.49\textwidth}{!}{\includegraphics{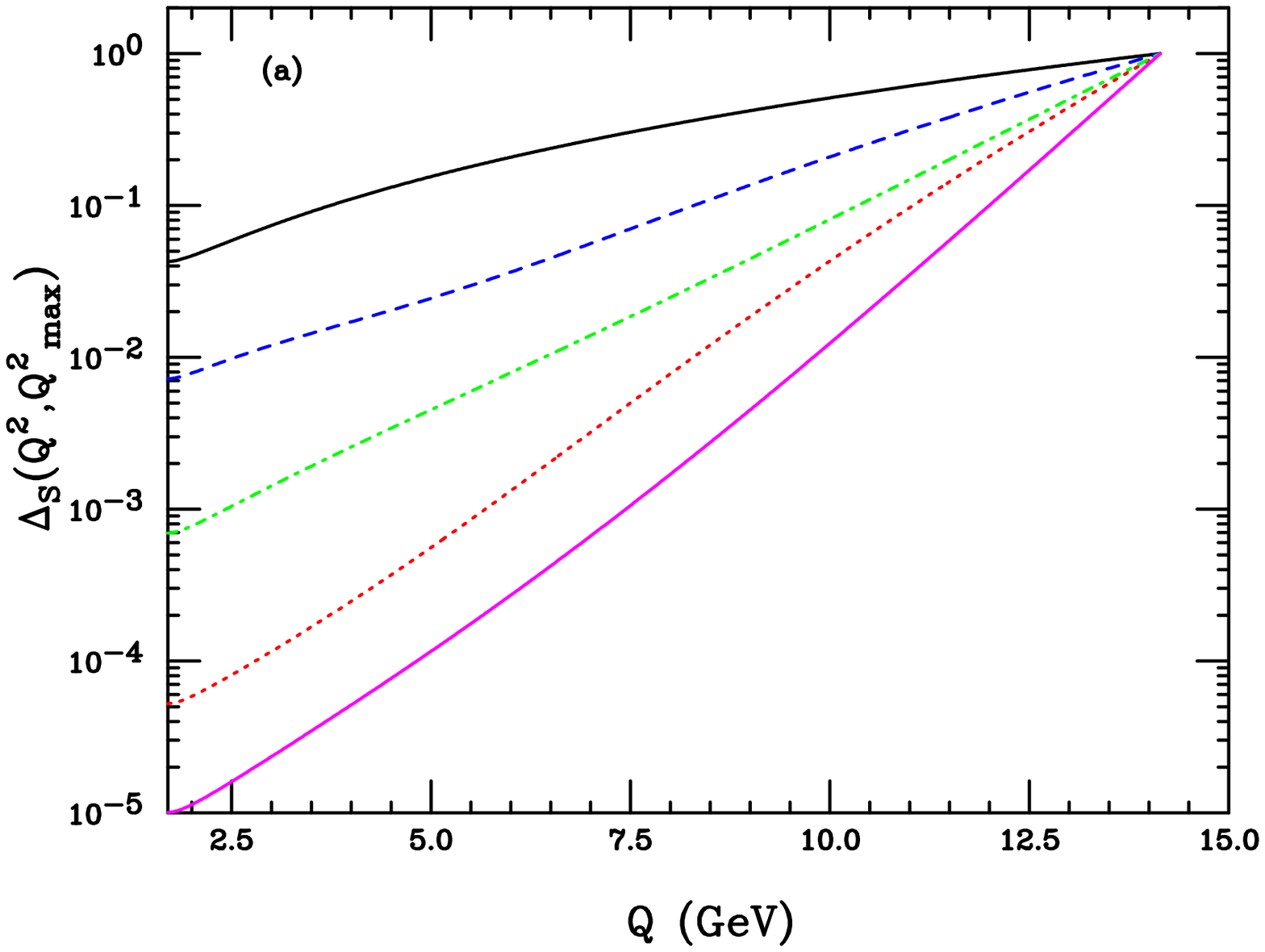}}%
\hfill%
\resizebox{0.49\textwidth}{!}{\includegraphics{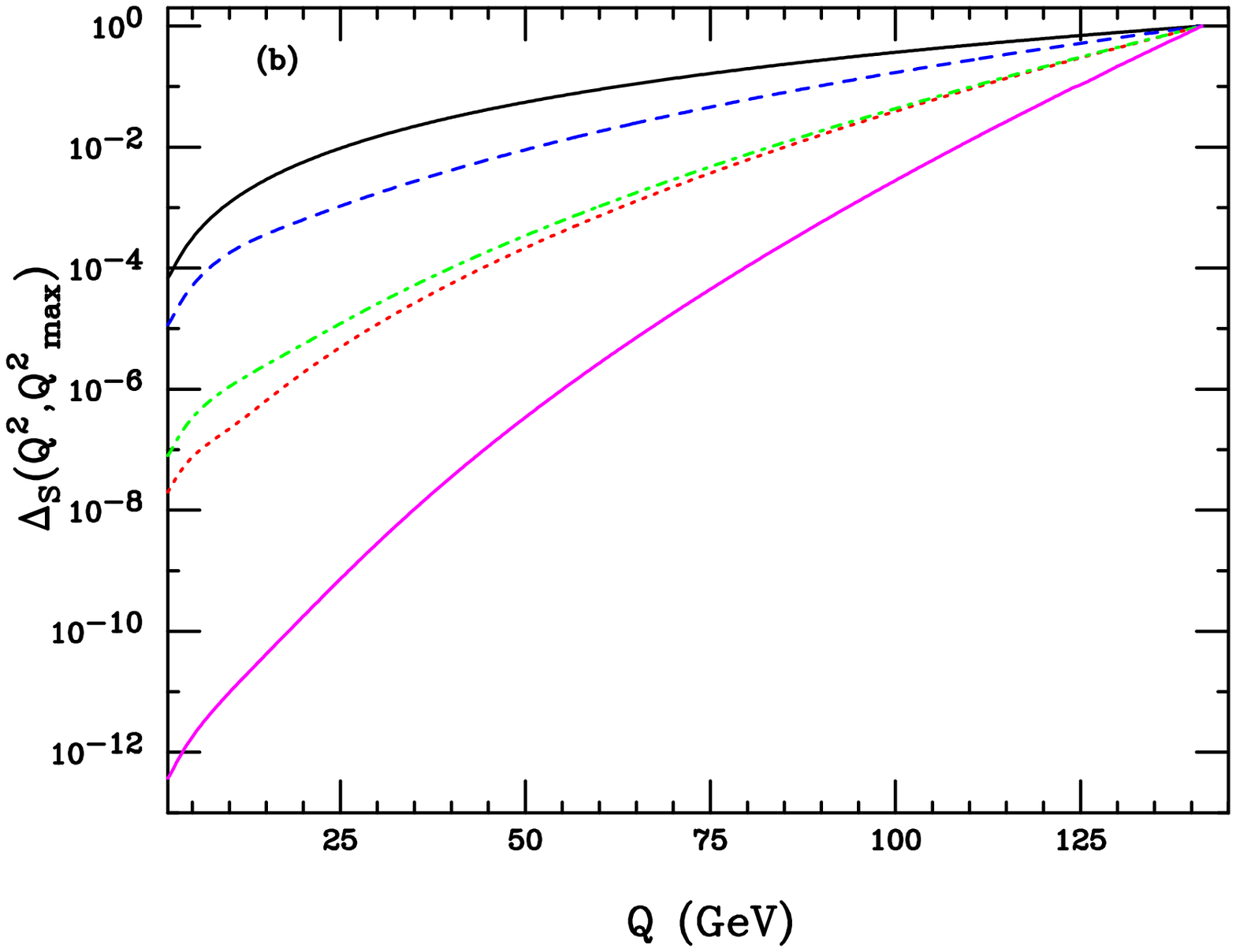}}}
\caption{Gluon Sudakov form factors, in the vacuum (solid, black) and
in media with $\hat qL_0=2$ (dashes, blue), 5
(dots, red), 20 (dot-dashes, green) and 50 (magenta, solid) GeV$^2$.
The shown cases correspond to starting energies of $E=10$~GeV (a)
and 100 GeV (b).
(For the interpretation of the references to colours in all figure
legends, the reader is referred to the online version of this paper).} 
\label{sud}
\end{figure}

In Fig.~\ref{sud} we present the HERWIG Sudakov form factor,
possibly including medium effects by means of Eq.~(\ref{dp});
for simplicity, we plot $\Delta(Q^2,Q^2_{\mathrm{max}})$
with fixed $E$ and $L=L_0$, as if we were dealing with the 
the first emission in the shower.
We consider medium-induced showers
initiated by gluons of energy $E=10$ and 100 GeV and media with
$\hat q=$~1 and 10~GeV$^2$/fm, $L_0=2$ and 5~fm.
In the following, for the sake of brevity,
we shall label such medium configurations
in terms of the accumulated transverse momentum, say $\hat qL_0=2$,
5, 20 and 50 GeV$^2$; the corresponding values of $\hat q$ and $L_0$
are left understood.
For gluon-initiated showers, the lower value of the
evolution variable is, by default, $Q_0\simeq 1.7$~GeV \cite{herwig},
whereas we set by hand the upper value\footnote{The maximum value
for the virtuality-ordered PYTHIA showers used in \cite{acs} is 
$p_{\mathrm{max}}^2=4E^2$. Hence, even the $p^2$-evolution range in PYTHIA
is larger than the $Q^2$ one in HERWIG.} $Q_{\mathrm{max}}=\sqrt{2}~E$.
\par
From Fig.~\ref{sud} we learn that 
medium-induced effects are pretty relevant: for both $E$=10 and 100 GeV, the
suppression due to including $\Delta P(z,p^2,E,\hat q,L)$
can be up to several orders of magnitude, especially for relatively
small values of $Q$.
Of course, when
$Q$ approaches $Q_{\mathrm{max}}$, all form factors tend to 1, and
therefore the discrepancy due to the different values of $\hat qL_0$ tends
to become smaller, but it is nonetheless well visible
throughout all $Q$-range.
As for the behaviour of the modified $\Delta_S(Q^2,Q^2_{\mathrm{max}})$ when
$\hat q$ and $L_0$ change, we find that, for both $E$ values and in all 
$Q$-range, the highest 
no-branching probability is obtained for $\hat qL_0=2$~GeV$^2$ and the lowest
for $\hat qL_0=50$~GeV$^2$, as one should indeed expect.
The comparison between the other two options in instead more cumbersome:
for $E=10$~GeV, $\Delta_S(Q^2,Q_{\mathrm{max}})$ 
for $\hat qL_0=20$~GeV$^2$ is much larger than the form factor for
 $\hat qL_0=5$~GeV$^2$, thus implying more branching in the case of
accumulated transverse momentum equal to 5~GeV$^2$. 
For $E=$~100 GeV, the $\hat qL_0=20$~GeV$^2$ form factor is above
the $\hat qL_0=5$~GeV$^2$ one at low $Q$, but for $Q$ values larger than
70-80~GeV, which roughly
correspond to the first branchings in the shower, they are very close
to each other.\footnote{Note that the radiative energy loss is not  
only function of the product $\hat qL_0$, but its dependence on $\hat q$ and
$L_0$ is more complex.} 
Therefore, we can already foresee some similarities
between these two medium parametrizations in the phenomenological
analysis which we shall carry out for $E=100$~GeV.
In any case, the splitting-parton energy and the medium length
vary throughout the cascade and Fig.~\ref{sud} has been instead
obtained for fixed $L_0$ and $E$; the plotted 
$\Delta_S(Q^2,Q^2_{\mathrm{max}})$,
though very useful to give us a qualitative estimate of the
medium suppression,
are not exactly the ones which we shall implement 
within the HERWIG algorithm.

Since the ordering variables in PYTHIA and HERWIG are different,
the numerical comparison between Fig.~\ref{sud} and the
Sudakov form factors presented in \cite{mmff} is not straightforward.
However, as we said before, since the evolution range for variables
$z$ and $Q^2$ in HERWIG is narrower than for the corresponding
PYTHIA quantities, at least in the default scenarios,
we can already predict a stronger medium-induced Sudakov suppression
and more branchings in PYTHIA rather than in HERWIG, in both
vacuum and medium-modified showers.


\section{Results}

In this section we will present some
phenomenological results showing the impact of
medium modifications in HERWIG. In \cite{acs}, as a case study, the authors  
considered a gluon with fixed energy, giving rise to
a medium-modified cascade within the PYTHIA model.
This is not a standard option in HERWIG, but nevertheless, for the
sake of comparison, we implemented a fictitious process with
a single gluon of given energy $E$ initiating a shower.
In PYTHIA virtuality evolution, this was relatively straightforward,
whereas in HERWIG, since one has angular ordering, 
in principle one would need two colour-connected partons to
define the initial value of the evolution variable $Q^2_{\mathrm{max}}$.
Still, we managed to start with a gluon of energy 
$E$, setting by hand the upper value 
of the evolution variable $Q^2_{\mathrm{max}}=2E^2$.
Subsequent splittings ($g\to gg$ and so forth)
will follow standard colour coherence
\footnote{As an alternative option, one can run the
fictitious process $e^+e^-\to gg$ at $\sqrt{s}=2E$, 
available in the HERWIG library,
and study the parton shower only in one hemisphere.}.
Similar tricks would be necessary if one wanted to simulate such a fictitious
process in PYTHIA, but with showers ordered in transverse momentum, as their
implementation follows a dipole formalism and one would need to 
start with two colour-connected partons \cite{pythiakt}.
Furthermore, as we are not investigating an actual physical process,
we turn matrix-element corrections off; how to implement 
exact higher-order corrections to the hard-scattering
process in a medium is anyway
an open issue.

As in \cite{acs}, we shall study the distributions of the 
transverse momentum $p_T$ of final-state partons, the emission angle $\theta$,
with respect to the initial-gluon axis, and  the
logarithmic energy fraction, defined as  $\xi=\ln(E/|p|)$, 
$|p|$ being the modulus of the parton momentum.
As we said before, we consider showers in the vacuum and in a
dense medium characterized by parameters $\hat q$ and
$L_0$. We shall assume gluon energies $E=10$ and 100~GeV
and, as in the previous section, medium parameters leading to
$\hat qL_0= 2$, 5, 20 and 50 GeV$^2$. Throughout the cascade, 
the effective medium length $L$ will be obtained
by applying Eq.~(\ref{ell}); our distributions will be labelled
in terms of the accumulated transverse momentum
$\hat q L_0$.

Though being an unphysical quantity, it is interesting to
compute first the average parton multiplicity $\langle N\rangle$ 
in the vacuum and in media 
characterized by the parameters given above. Such multiplicities strongly
depend on the shower cutoff $Q_0$ and obviously increase whenever
$Q_0$ is lowered. In Table~\ref{mult} we quote these numbers for default
HERWIG and different values of
$\hat qL_0$: as 
expected, parton multiplicities increase in a dense medium, 
which is in agreement
with the measurements of larger medium-induced energy loss.
For $E=10$~GeV, the enhancement runs from 20\% ($\hat qL_0=2$~GeV$^2$) 
up to about 80\% ($\hat qL_0=50$~GeV$^2$); for $E=100$~GeV the
corresponding numbers are 7\% and 70\%.
It is interesting to notice in Table~\ref{mult} that the cases 
$\hat qL_0=5$ and 20~GeV$^2$ give quite similar results and that 
the $\hat qL_0=20$~GeV$^2$
multiplicity is slightly above the $\hat qL_0=5$~GeV$^2$ 
one for $E=10$~GeV
and below for a gluon of 100 GeV. In fact, the modified splitting
functions (\ref{med}) do not scale with respect to $\hat qL_0$ and do
depend on the initiating parton energy: therefore, the phenomenological
implications of the introduction of medium effects depend on the three
variables $\hat q$, $L_0$ and $E$ and different behaviours are
to be expected if we start the cascade with partons of different energies,
even though we use same values of $\hat q$ and $L_0$.
In Fig.~(\ref{sud}), although the Sudakov form factors were
computed for fixed $E$ and $L$, we already noticed differences
between the options $\hat qL_0=5$~GeV$^2$ 
and $\hat qL_0=20$~GeV$^2$, according to
whether we have $E=10$ or 100 GeV.
\begin{table}
\small
\caption{\label{mult}
Average parton multiplicities in HERWIG showers initiated by gluons 
of energy of 10 and
100 GeV, in the vacuum and in a medium with assigned values of $\hat qL_0$.}
\begin{center}
\begin{tabular}{| c || c | c | c | c | c ||}
\hline
$E$ & $\hat q L_0=0$ & 
$\hat qL_0=2$~GeV$^2$ & $\hat qL_0=5$~GeV$^2$ & 
$\hat qL_0=20$~GeV$^2$ & $\hat qL_0=50$~GeV$^2$ \\
\hline 
10~GeV & 2.56 & 3.05 & 4.14 & 3.60 & 4.56\\
\hline
100~GeV & 6.95 & 7.41 & 8.79 & 8.93 & 11.70\\
\hline
\end{tabular}
\end{center}
\end{table}\par
\begin{figure}[t]
\centerline{\resizebox{0.49\textwidth}{!}{\includegraphics{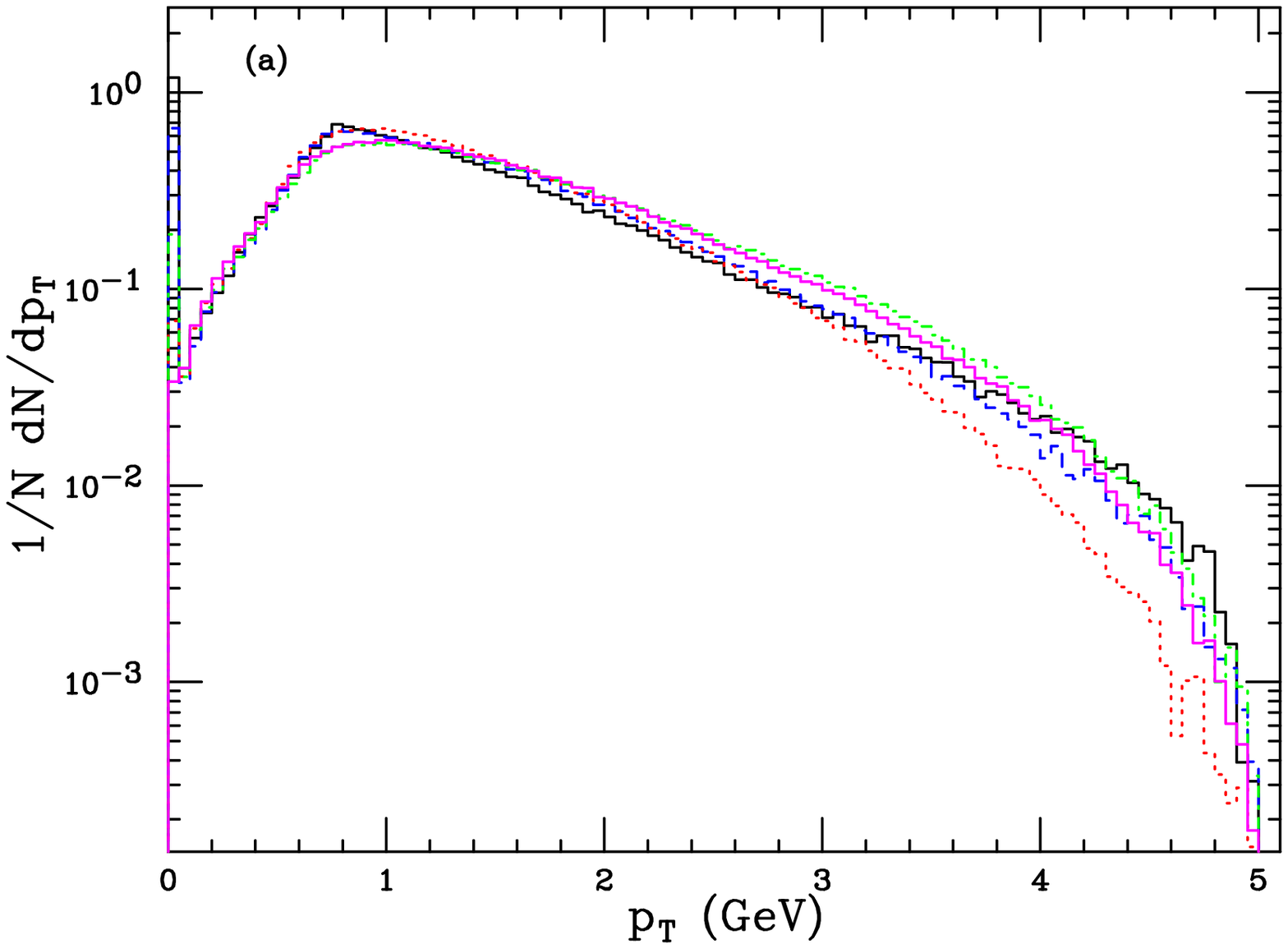}}%
\hfill%
\resizebox{0.49\textwidth}{!}{\includegraphics{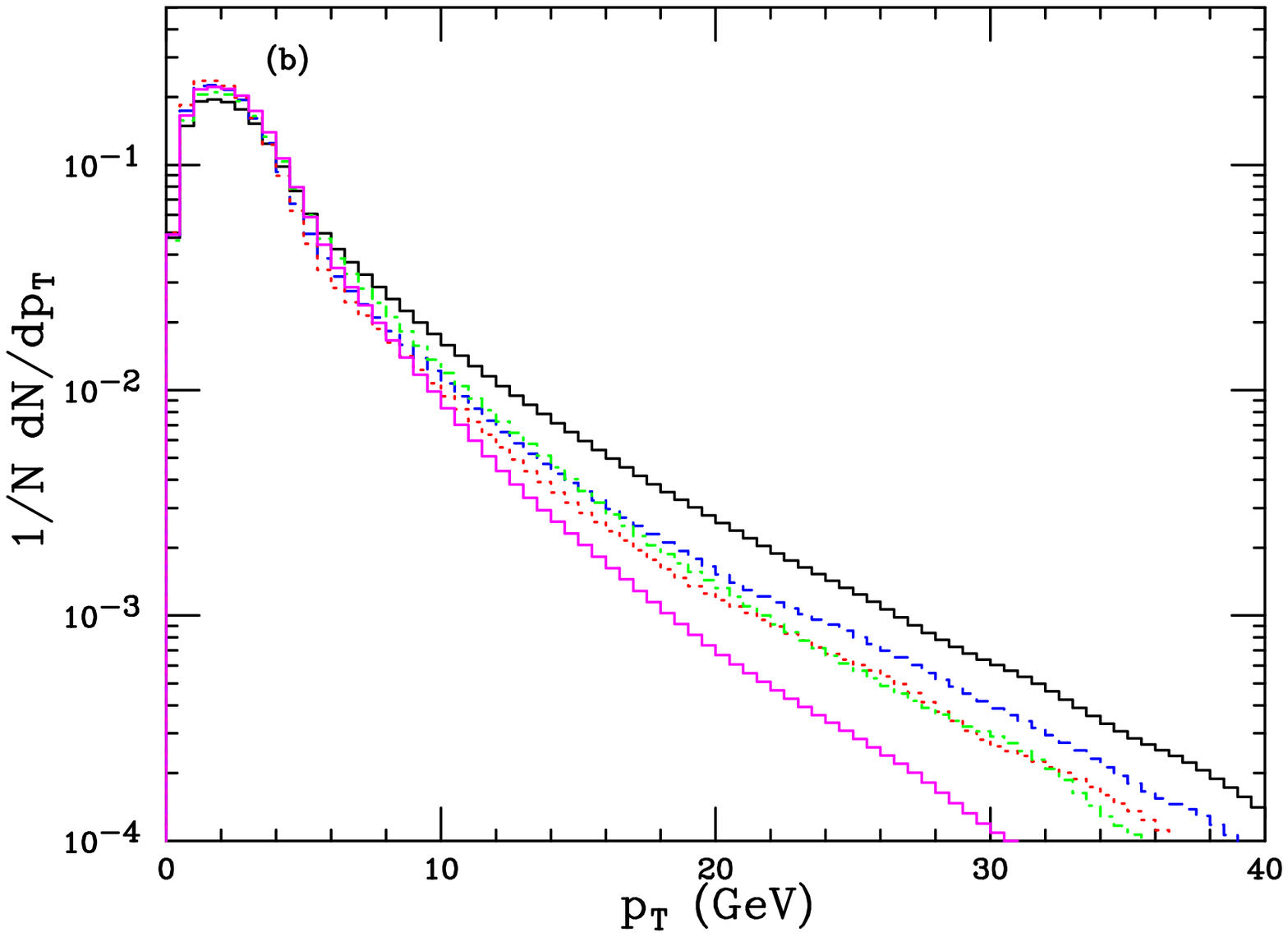}}}
\caption{Transverse momentum multiplicity for a medium-modified 
shower initiated by a gluon of energy 10 GeV (a) and 100 GeV (b),
in the vacuum (solid, black) and in media with accumulated
transverse momentum $\hat qL_0=2$ (dashes, blue), 
5 (dots, red), 20 (dot-dashes, green) and 50 (solid, magenta) GeV$^2$.}
\label{pt}
\end{figure}

\begin{figure}[ht!]
\centerline{\resizebox{0.49\textwidth}{!}{\includegraphics{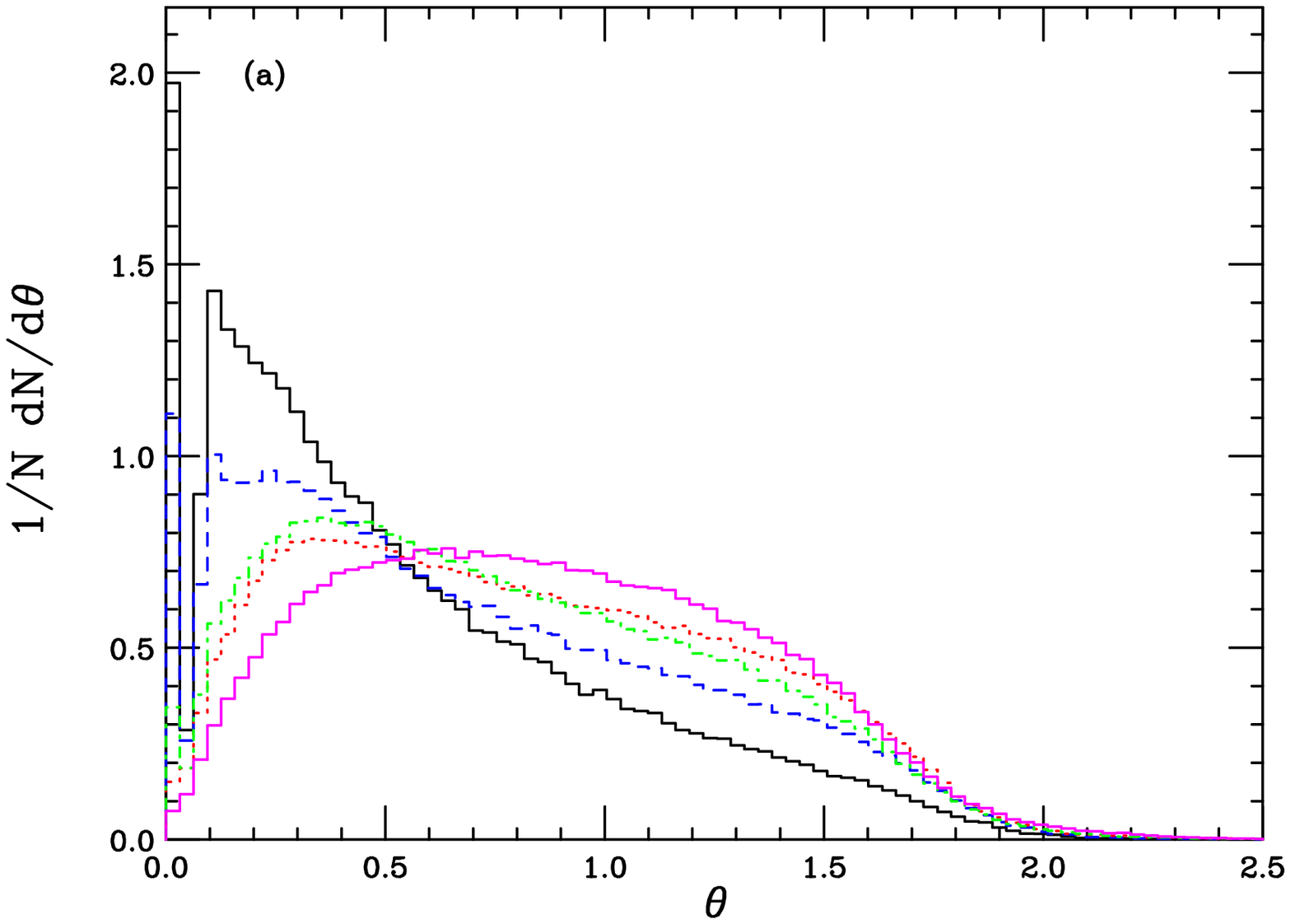}}%
\hfill%
\resizebox{0.49\textwidth}{!}{\includegraphics{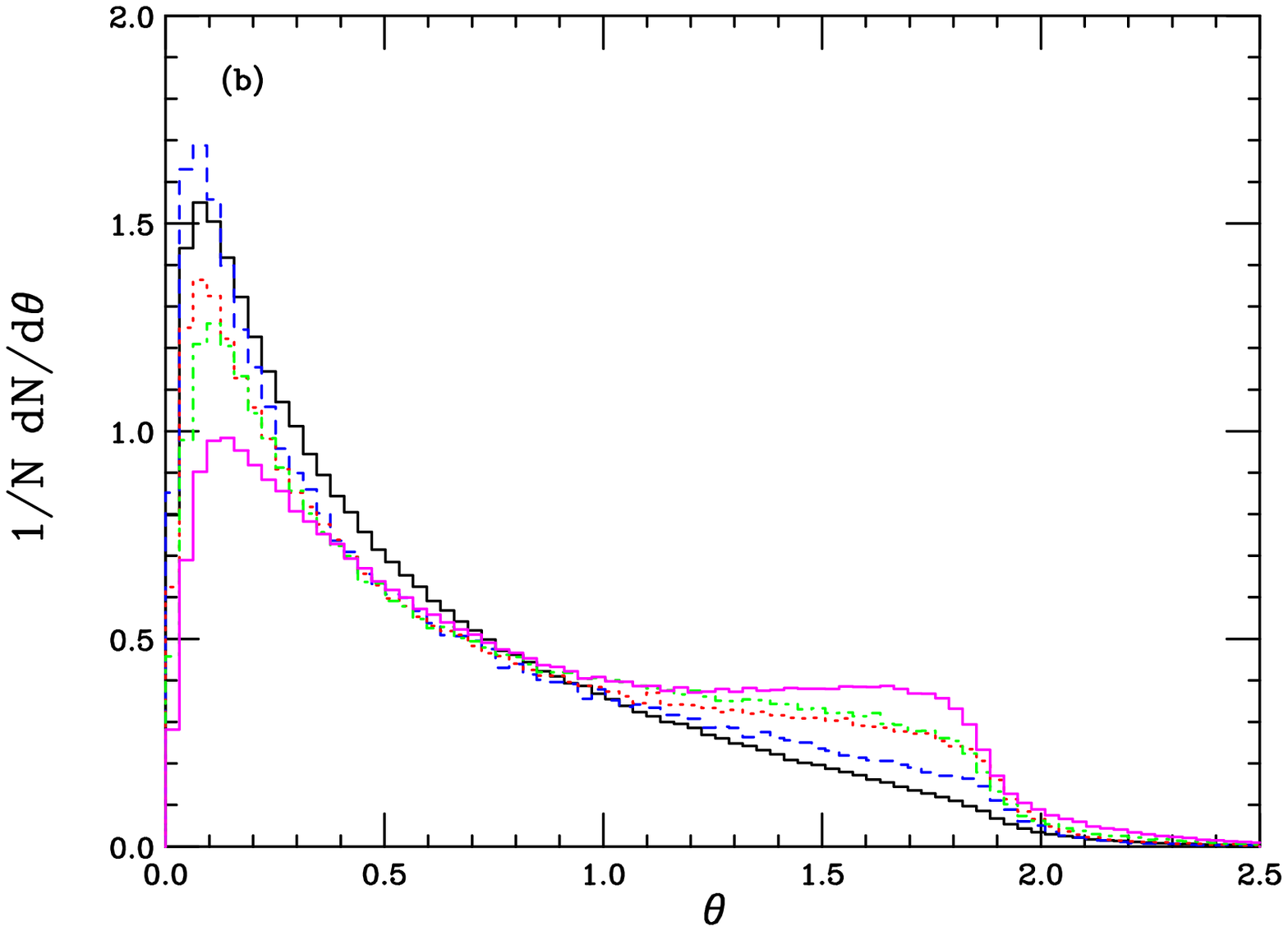}}}
\caption{As in Fig.~\ref{pt}, but showing the
angular distributions for showers 
initiated by a gluon of energy 10 GeV (a) and 100 GeV (b),
in the vacuum and in a dense medium.}
\label{teta}
\end{figure}

\begin{figure}[ht!]
\centerline{\resizebox{0.49\textwidth}{!}{\includegraphics{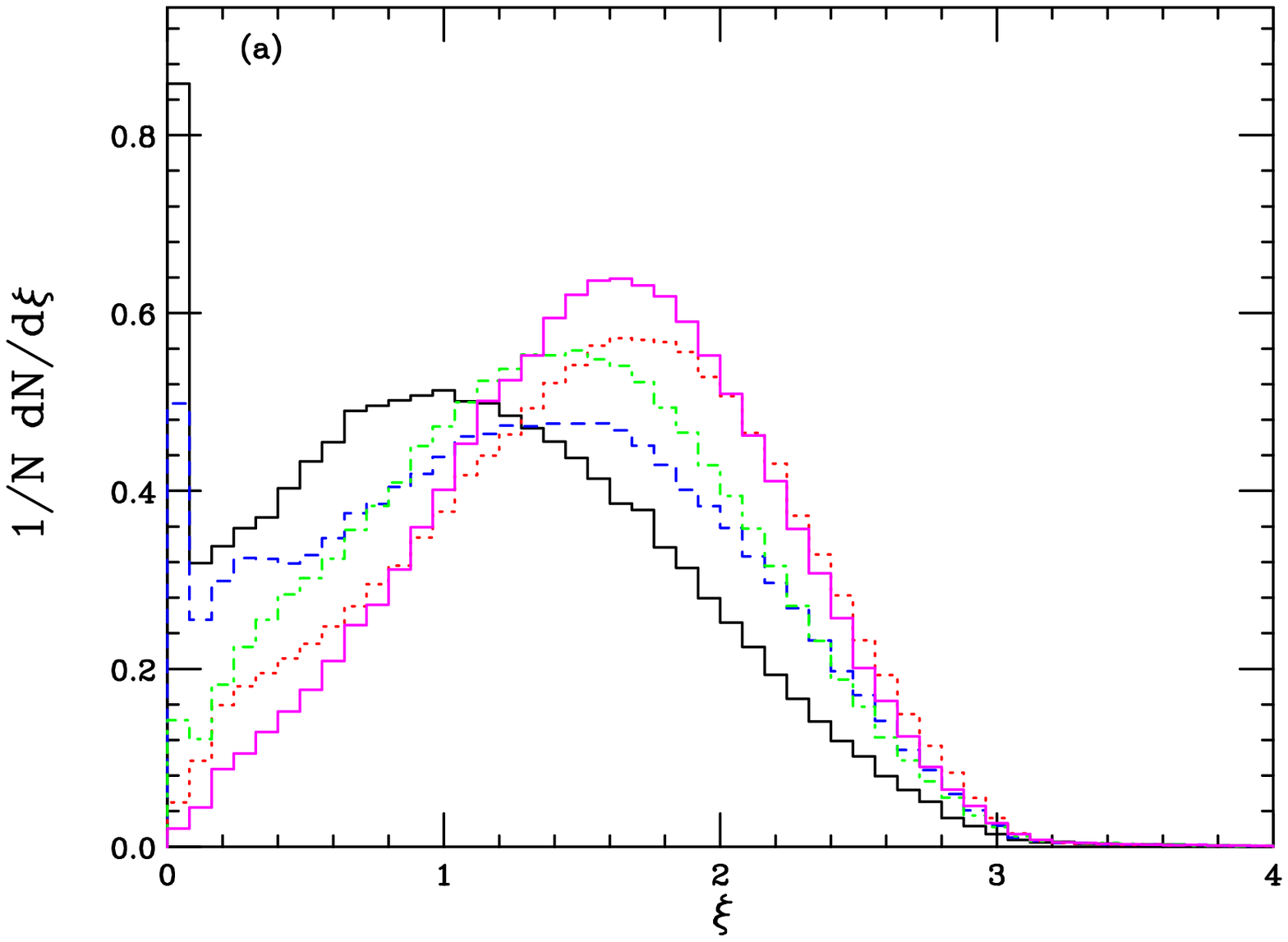}}%
\hfill%
\resizebox{0.49\textwidth}{!}{\includegraphics{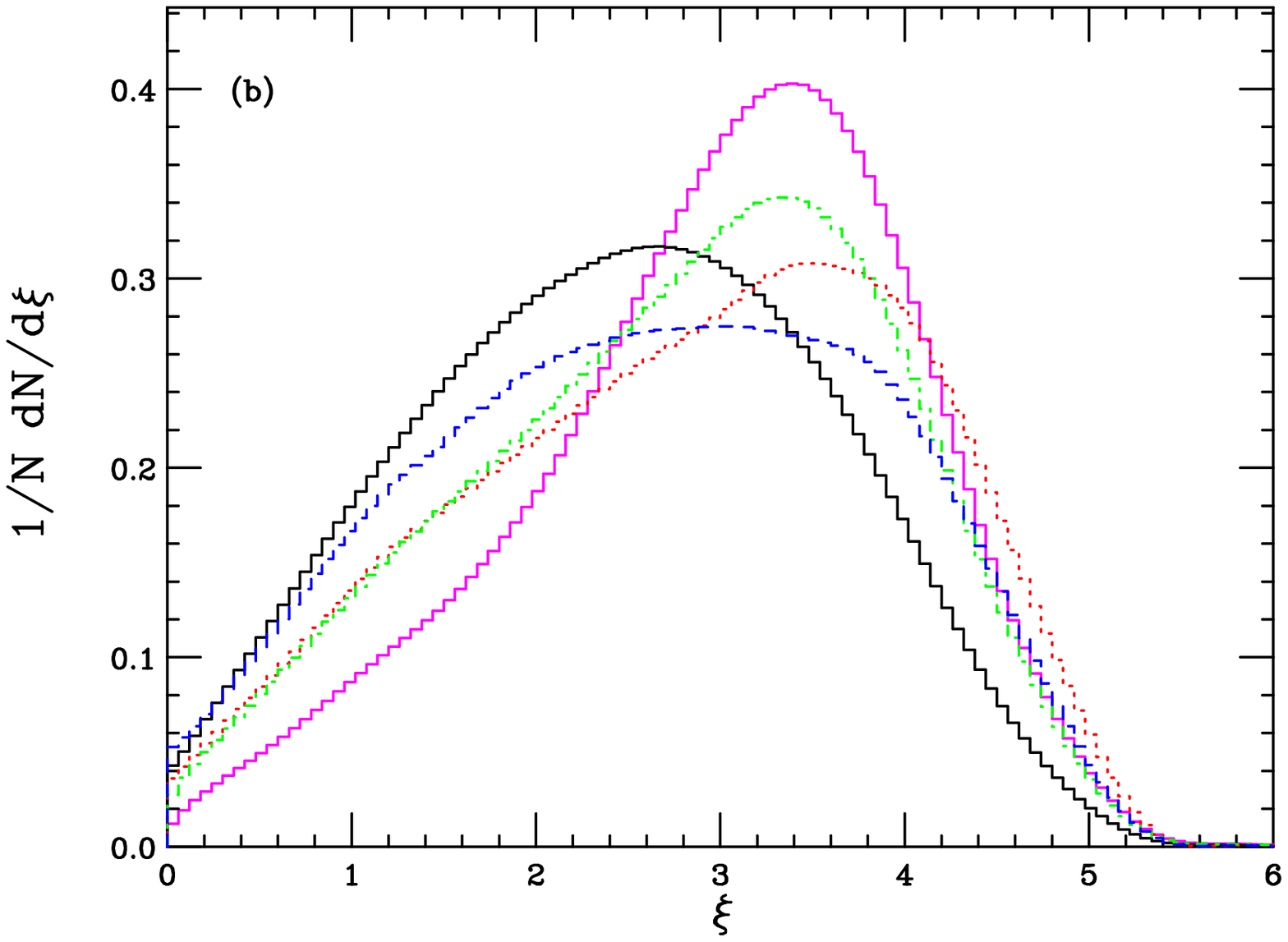}}}
\caption{As in Figs.~\ref{pt} and \ref{teta}, but presenting
the logarithmic energy-fraction ($\xi$).}
\label{xi}
\end{figure}

  
In Figs.~\ref{pt}--\ref{xi} we present the $p_T$, 
$\theta$ and $\xi$ distributions, according to HERWIG in the vacuum and
in a medium with the parameters $\hat q$ and $L_0$ stated before, for
energies $E=10$~GeV (a) and 100 GeV (b).
We plot everywhere normalized parton multiplicities, such as
$(1/N)\  dN/dp_T$, so that the total integral will be equal to 
1.\footnote{On the contrary, the plots presented in Ref.~\cite{acs} are
normalized to the mean parton multiplicity; rescaling our
distributions to the multiplicities in Table~\ref{mult} is straightforward.}
A feature of the vacuum spectra at 10~GeV is that they exhibit a sharp peak
at $p_T=\xi=\theta=0$, corresponding to events where 
the initiating gluon evolves down to the shower cutoff with 
no branching at all. The probability for this to happen depends on
the Sudakov form factor $\Delta_S(Q^2_{\mathrm{max}},Q^2_0)$, 
hence it may
be eventually lowered by decreasing $Q_0$ or increasing $Q_{\mathrm{max}}$,
thus allowing more radiation in the evolution. 
Also, such a spike will vanish once hadronization is turned on.
A peak, though less sharp, is visible even for media with small values
of $\hat qL_0$, e.g. $\hat qL_0=2$~GeV$^2$, 
whereas it disappears for larger values of 
$\hat qL_0$. Also, it is less visible for $E=100$~QeV, as in this case
$Q_{\mathrm{max}}\simeq 141.42$~GeV and $\Delta(Q^2_{\mathrm{max}},Q^2_0)$
is a pretty small non-emission probability.\footnote{A 
similar effect was found, e.g., in Ref.~\cite{corsey}, where the simulation
of $W/Z$ transverse momentum at the Tevatron exhibited a 
sharp peak at
$p_T=0$, which disappears whenever one sets 
a non-zero intrinsic transverse momentum
for the incoming partons.} It is interesting to notice that,
even for $E=10$~GeV,
such a peak was not so visible in \cite{acs}, where the authors did find
some events with no radiation, but, since the branching probability in the
PYTHIA model is higher that in HERWIG in all evolution range,  
the Q-PYTHIA spectra exhibit smooth behaviour, even when $p_T$, $\theta$ or
$\xi$ approach zero. 
Later in this section, as an example, we shall plot integrated
parton multiplicities, which behave smoothly also in the very first bin
of our histograms.
\par
As for the $p_T$ spectra, i.e. Fig.~\ref{pt}, 
for $E$=10~GeV the vacuum distribution is 
above the others at large $p_T$, as in \cite{acs}; however,
due to the presence of the $p_T=0$ peak, such a behaviour is less 
evident than in PYTHIA. In both cases, it is anyway the $\hat qL_0=5$~GeV$^2$
curve the one that gives the lowest normalized multiplicity at large $p_T$.
At low $p_T$, Q-PYTHIA gives the highest parton
multiplicity in the two scenarios corresponding to $L=5$~fm; the
lowest multiplicity  was instead the one for the vacuum case \cite{acs}.
In HERWIG the low-$p_T$ distribution is clearly affected by the
$p_T\to 0$ behaviour: the spectra exhibit a maximum about $p_T=1$~GeV, but, 
up to $p_T\simeq 0.8$~GeV, the difference between the four
chosen $\hat qL_0$ is quite small.
For $E=100$~GeV, 
the absence of a significant event fraction
with no showering makes it
easier to compare with the results yielded by Q-PYTHIA.
Overall, for the reasons discussed above, namely lower effects in
the Sudakov form factor, the impact of the implementation of
medium modifications in HERWIG looks smaller than in Q-PYTHIA.
However, the behaviour with respect to varying $\hat q$ and $L_0$ is
qualitatively the same.
The vacuum spectrum is the lowest at small $p_T$
and the highest at large $p_T$, where the 
$\hat qL_0=50$~GeV$^2$ distribution is the one going to zero more rapidly.
The other three curves, corresponding to $\hat qL_0$=2, 5 and 20~GeV$^2$, 
lie in the range of the two extreme cases, as one should expect.
In particular, at large $p_T$, $\hat qL_0= 5$ and 20~GeV$^2$ 
seem to give similar
distributions, whereas the $\hat qL_0= 2$~GeV$^2$ option yields the largest
high-$p_T$ normalized multiplicity after the vacuum.

Concerning the $\theta$ spectra, i.e. Fig.~\ref{teta},
first of all we need to point out an essential
difference between HERWIG and PYTHIA angular distributions, 
independently of medium effects. PYTHIA, which does not
systematically implement angular ordering, allows parton radiation in all 
allowed phase space, 
i.e. up to $\theta=\pi$; on the contrary, in HERWIG 
angular-ordered showers,
radiation is possible only for $\zeta<1$, $\zeta$ being the
showering variable defined in Section 2.
This leads to an empty region in the physical phase space,
corresponding to large-angle emission.
More precisely, 
in the massless approximation, the condition $\zeta<1$ implies
$\theta<\pi/2$ in the HERWIG showering frame. At the end of the
showering, however, when jets acquire mass, a Lorentz boost is applied:
this enlarges the angular permitted region, but a zone where radiation
is forbidden is still present. In a physical process, such as 
$e^+e^-\to q\bar q$, this empty region would be filled by the soft
radiation from the other jet, i.e. from $\bar q$ if one triggers the
shower initiated by $q$ \footnote{In any case, even with two jets,
there will still be a missing region,
due to hard and wide-angle radiation, that one should eventually fill 
by means of matrix-element corrections.}.
However, as we are simulating an unphysical process with only a single
coloured parton, a dead zone, corresponding to large-angle radiation
off the primary gluon, must be expected.

In fact, the spectra in Fig.~\ref{teta} show that the parton
multiplicity above $\theta\simeq 2$ is negligible. As medium modifications
mostly affect the first few emissions, which, because of angular
ordering, are the ones at the
largest $\theta$, it is reasonable
that the spectra in a dense medium are above the vacuum ones
in $1<\theta<2$, as we learn from Fig.~\ref{teta}.
At small $\theta$, i.e. far from the hard scattering,
the parton formation length $2zE/k_T^2$ is likely to be large and the
effective medium length $L$, given by Eq.~(\ref{ell}), becomes negative.
This implies that small-angle branchings are typically 
vacuum-like.
As far as the different $\hat qL_0$ options are concerned, 
the $\hat qL_0=2$~GeV$^2$ is
clearly the closest to the vacuum prediction, the difference between
$\hat q L_0=5$ and 20~GeV$^2$ is small, though visible, whereas the
$\hat q L_0=50$~GeV$^2$ spectrum, corresponding to the strongest medium
modifications, leads to the highest parton multiplicity in the
allowed large-angle range, say $1<\theta <1.8$. 
\par
We finally comment on 
the logarithmic energy-fraction ($\xi$) plots, presented in
Fig.~\ref{xi}. As for the $E=10$~GeV case, the $\xi=0$ peak clearly
spoils the low-$\xi$ part of the spectrum. However, regardless of such 
a peak, the suppression of 
partons with small energy fraction in a dense medium with respect to the
vacuum is still clear.
The behaviour of the four medium-modified spectra in terms of 
$\hat q$ and $L_0$ is alike the one displayed in \cite{acs}.
At small $\xi$, the $\hat qL_0=2$~GeV$^2$ 
spectrum yields the highest energy fraction,
followed, in order, by $\hat qL_0=20$, 5 and 50~GeV$^2$.
Around the peak, the order is reversed, with $\hat qL_0=50$~GeV$^2$ giving the
largest event fraction and $\hat qL_0=2$~GeV$^2$ the lowest.
At large $\xi$, the difference due to the different options of
$\hat qL_0$ tends to become smaller; at
very high $\xi$, the $\hat qL_0$=5~GeV$^2$ spectrum is above the others.
For $E=100$~GeV, since the
events without emissions are very few, the comparison among the different
curves and the results in \cite{acs} is straightforward.
The low-$\xi$ suppression in a medium with respect to the vacuum
is still remarkable: within the modified spectra, at small $\xi$, 
the $\hat qL_0=2$~GeV$^2$ option yields the highest event fraction,
followed by $\hat q L_0=5$ and 20~GeV$^2$, whose results are roughly
similar, and finally $\hat qL_0=50$~GeV$^2$. 
At large $\xi$, the order of the different $\hat qL_0$ cases is
reversed, as happens for $E$=10~GeV.

\begin{figure}[t]
\centerline{\resizebox{0.60\textwidth}{!}{\includegraphics{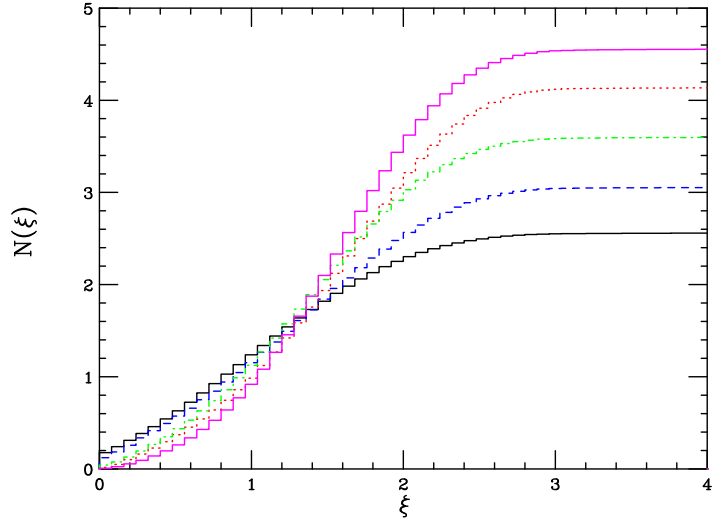}}}
\caption{Integrated $\xi$  distribution 
in the vacuum (solid, black) 
and in a medium with $\hat qL_0$= 2 (dashes, blue), 5 (dots, red), 
20 (dot-dashes, green) and 50 (solid, magenta) GeV$^2$.
The energy of the gluon initiating the parton shower has been set to 10 GeV.}
\label{xin}
\end{figure}

The presence of a peak for a gluon with $E=10$~GeV, whenever
$p_T$, $\xi$ or $\theta$ are close to zero,  can be 
a bit disturbing, since we have a non-negligible
fraction of events in the very first histogram bin, affecting
overall the full spectrum. 
As we said, a possibility to reduce 
the events in the first bin consists in enlarging the evolution range
or turning hadronization on.
As an alternative, for the time being, we can compute integrated quantities,
such as, e.g., parton multiplicities up to a given value of $\xi$,
namely: 
\begin{equation}
N(\xi)=\int_0^{\xi}{d\xi'}\frac{dN}{d\xi'}.\end{equation}
In Fig.~\ref{xin} we plot $N(\xi)$ in the
vacuum, for the four chosen scenarios of $\hat q$ and $L_0$,
and $E=10$~GeV: we observe a well visible medium-induced
suppression at small $\xi$ with respect to the vacuum.
At low energy fractions, e.g. $\xi\simeq 0.5$, we find that the
medium suppression is about 10\% for $\hat qL_0=2$~GeV$^2$
and can be up to 60\% for accumulated transverse momentum 
$\hat qL_0=50$~GeV$^2$. 
As we said in the introduction, although an actual comparison would
require hadronization corrections and possible tuning
of the Monte Carlo parameters to the data, such results are indeed 
quite encouraging, as they are in qualitative agreement with the
small-$\xi$ suppression observed at RHIC.

Before closing this section, following \cite{acs}, 
we wish to study the $p_T$, $\theta$ and $\xi$ distributions
for fixed values of the medium length throughout the parton cascade.
This means that, instead of applying Eq.~(\ref{ell}) at each branching,
we shall always employ $L=L_0$. In fact, because of Eq.~(\ref{ell}),
the medium length could even become negative at some point, 
which implies that, after the first few branchings, emissions 
occur in a vacuum-like fashion.
For $L=L_0$, therefore, 
the medium length will always stay positive,
and stronger medium effects are to be expected.
As a case study, we investigate the two options $\hat qL_0=2$ and
$\hat qL_0=50$~GeV$^2$, for variable and fixed medium length, and still
$E=10$ and 100 GeV. We first calculate
the average parton multiplicity $\langle N\rangle$: 
we obtain, for $L=L_0$ and $E=10$~GeV,
$\langle N\rangle\simeq 3.17$ 
and 4.92, for $\hat qL_0=2$ and 50 GeV$^2$, respectively.
This corresponds to enhancements of 4\% and 8\% with respect to the
variable-length results in Table~\ref{mult}.
For an initiating gluon of energy 100 GeV, the average multiplicities are
$\langle N\rangle\simeq$~8.09 ($\hat qL_0=2$~GeV$^2$) and 
$\langle N\rangle\simeq$~19.23 ($\hat qL_0=50$~GeV$^2$), hence
the enhancement runs from 10\% to 65\%.
Such results are in agreement with the expectation that a fixed
length should emphasize medium-induced effects.

\begin{figure}[ht!]
\centerline{\resizebox{0.49\textwidth}{!}{\includegraphics{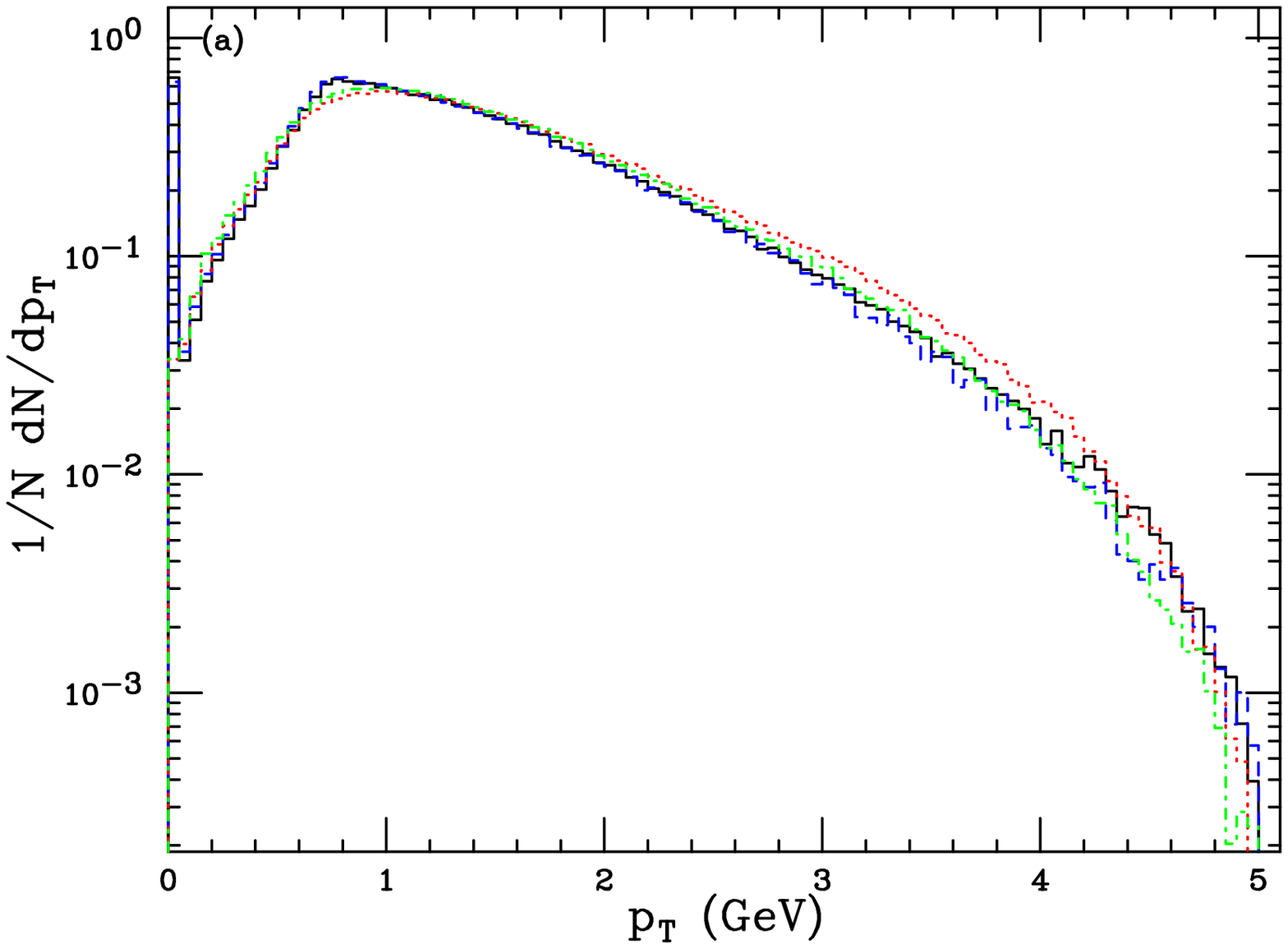}}%
\hfill%
\resizebox{0.49\textwidth}{!}{\includegraphics{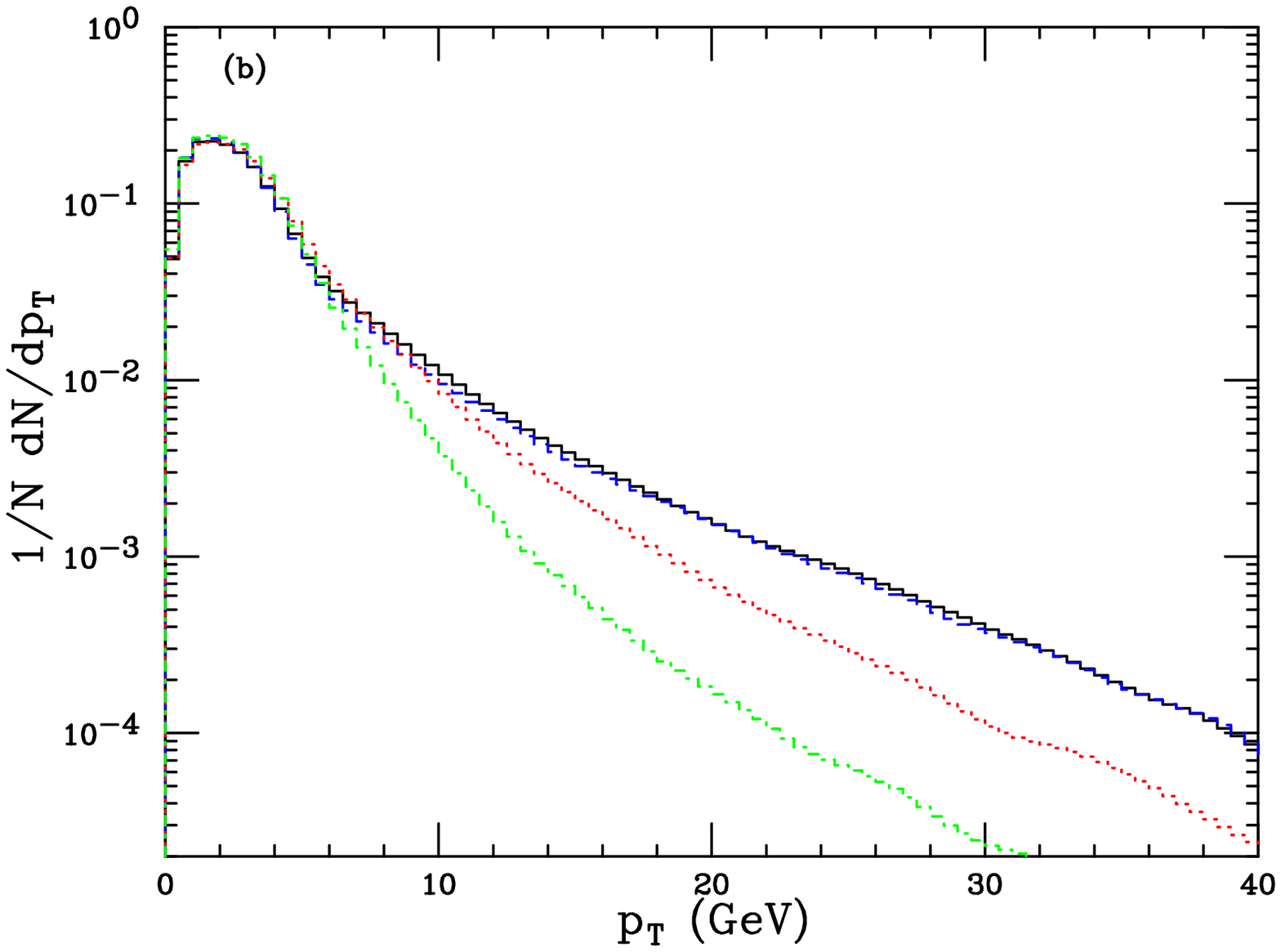}}}
\caption{Transverse momentum distribution,
with variable and fixed medium length  $L$.
The considered cases are $\hat qL_0=2$~GeV$^2$ (black, solid: variable $L$;
dashes, blue: fixed $L$)
and $\hat qL_0=50$~GeV$^2$ 
(red, dots: variable $L$; green, dot-dashes: fixed $L$),
with an initiating gluon of 10 GeV (a) and 100 GeV (b).}
\label{ptfix}
\end{figure}

\begin{figure}[ht!]
\centerline{\resizebox{0.49\textwidth}{!}{\includegraphics{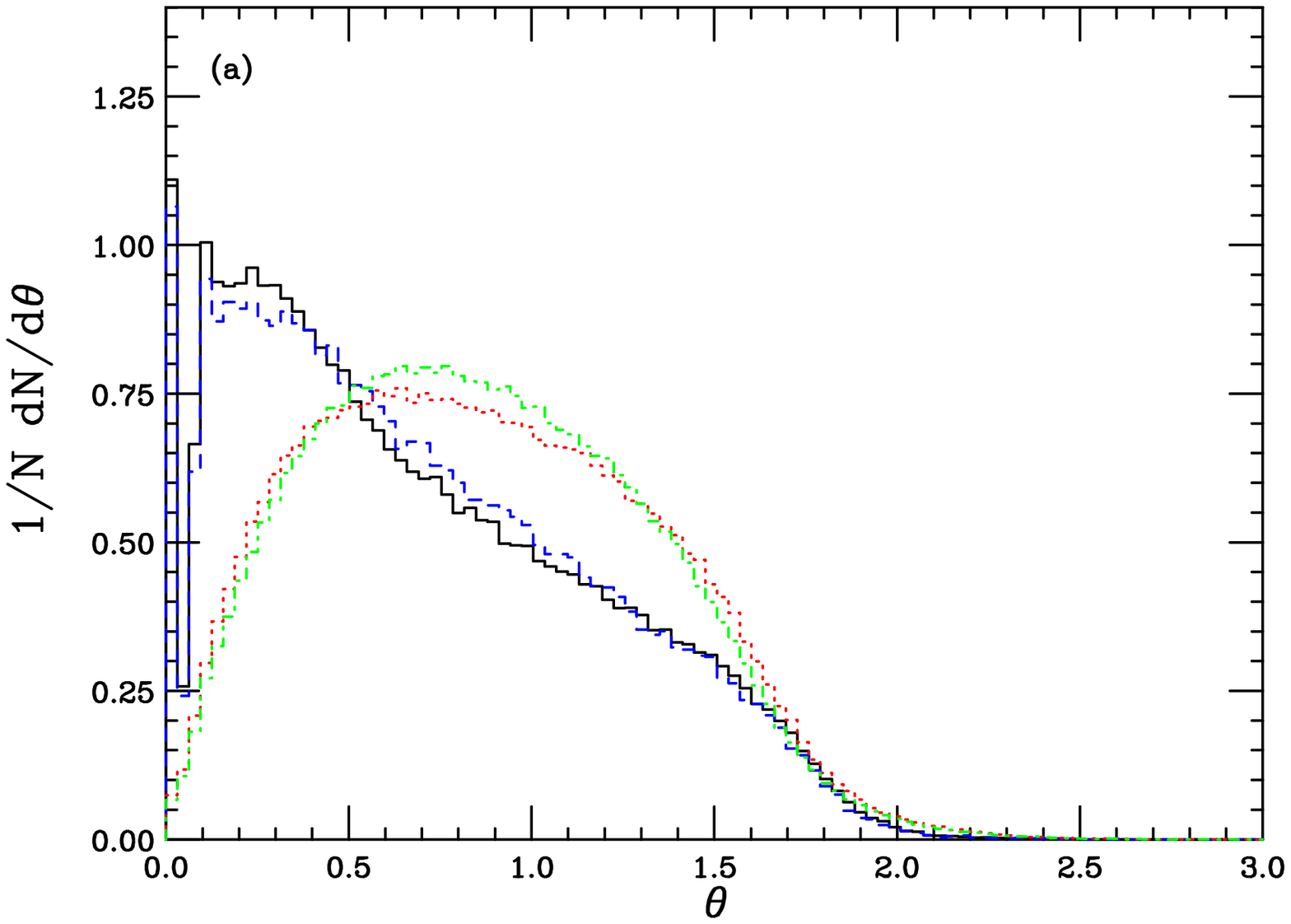}}%
\hfill%
\resizebox{0.49\textwidth}{!}{\includegraphics{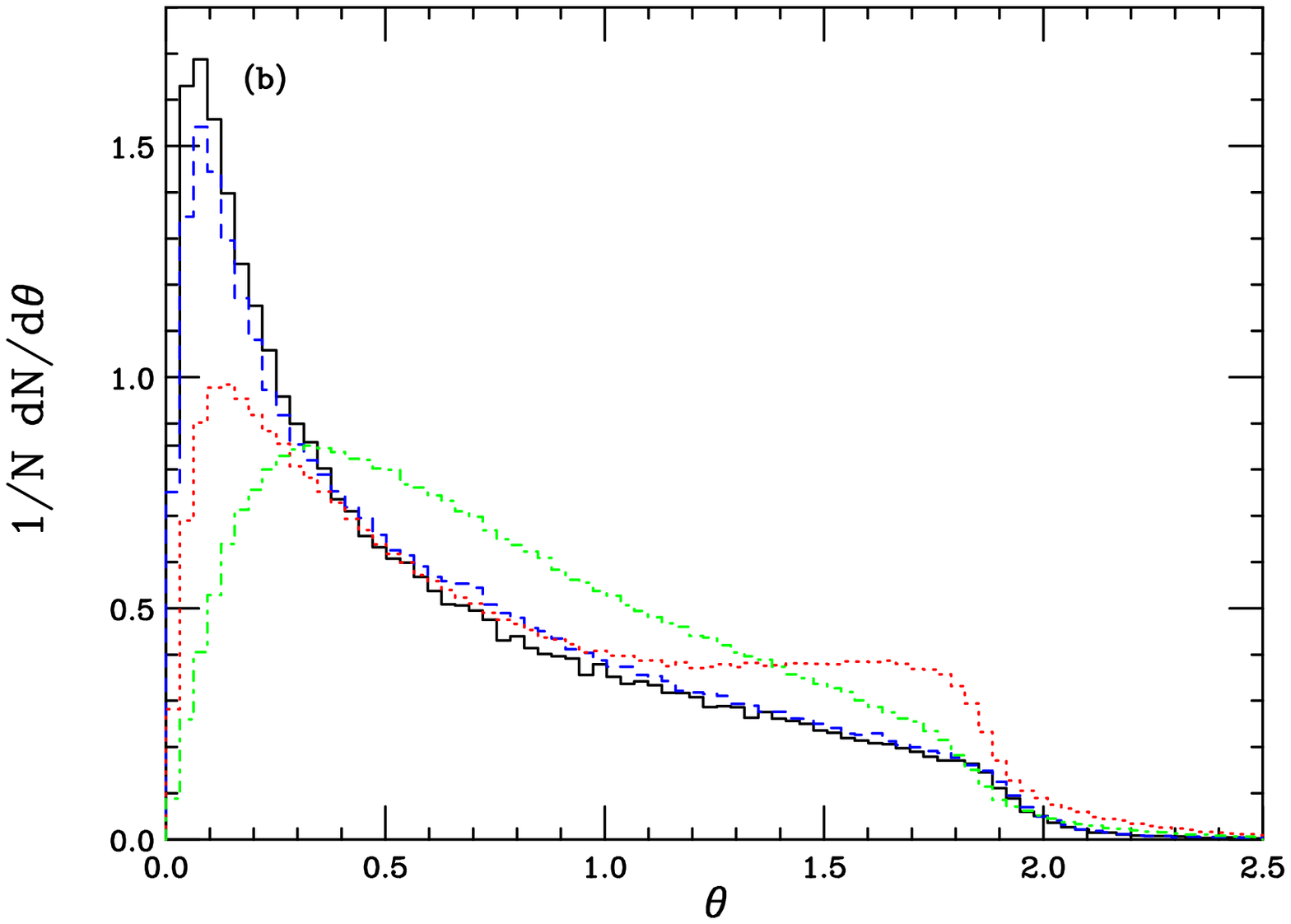}}}
\caption{As in Fig.~\ref{ptfix}, but displaying the
angular distributions for variable and fixed medium length.}
\label{tetafix}
\end{figure}

\begin{figure}[ht!]
\centerline{\resizebox{0.49\textwidth}{!}{\includegraphics{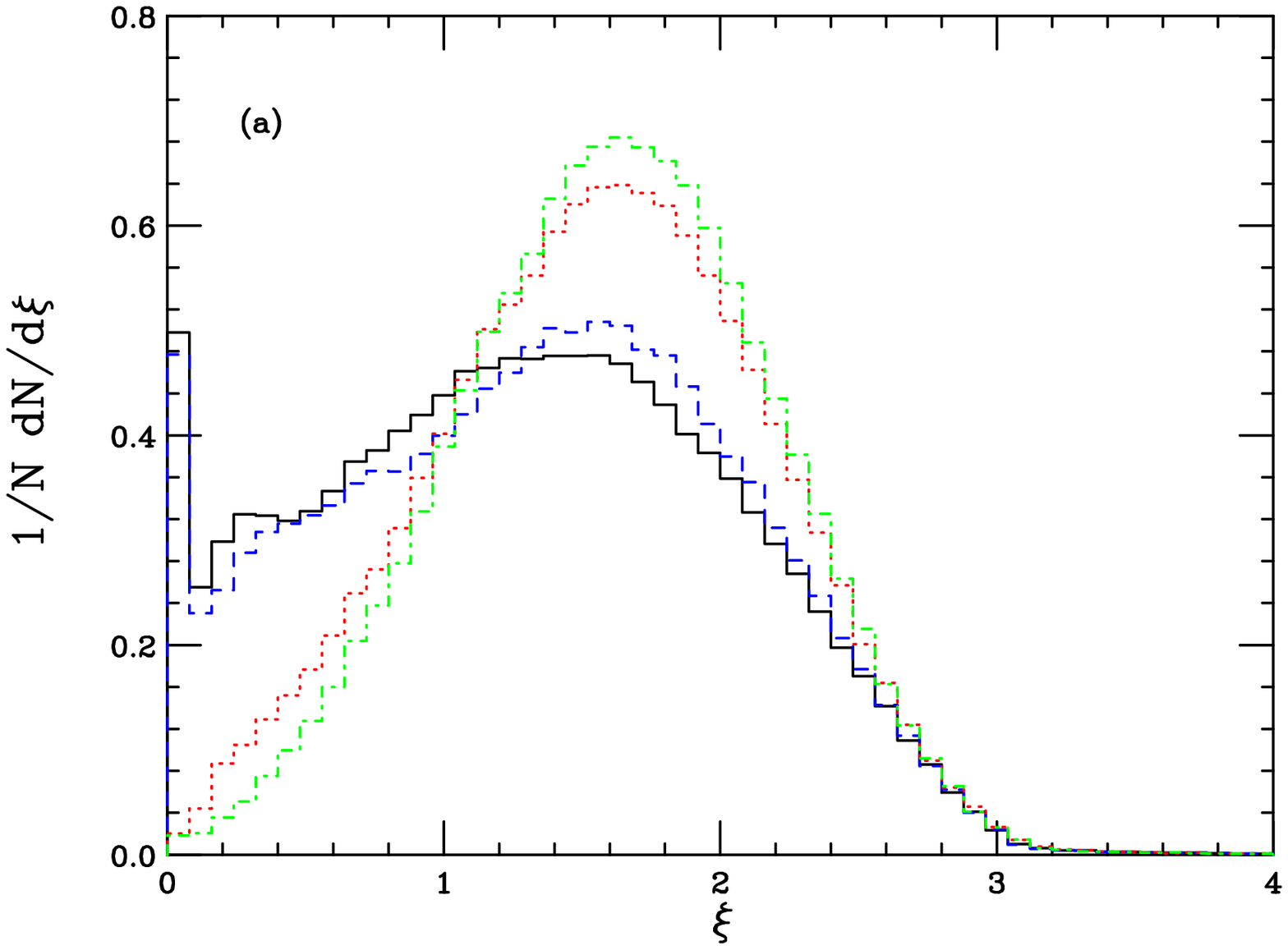}}%
\hfill%
\resizebox{0.49\textwidth}{!}{\includegraphics{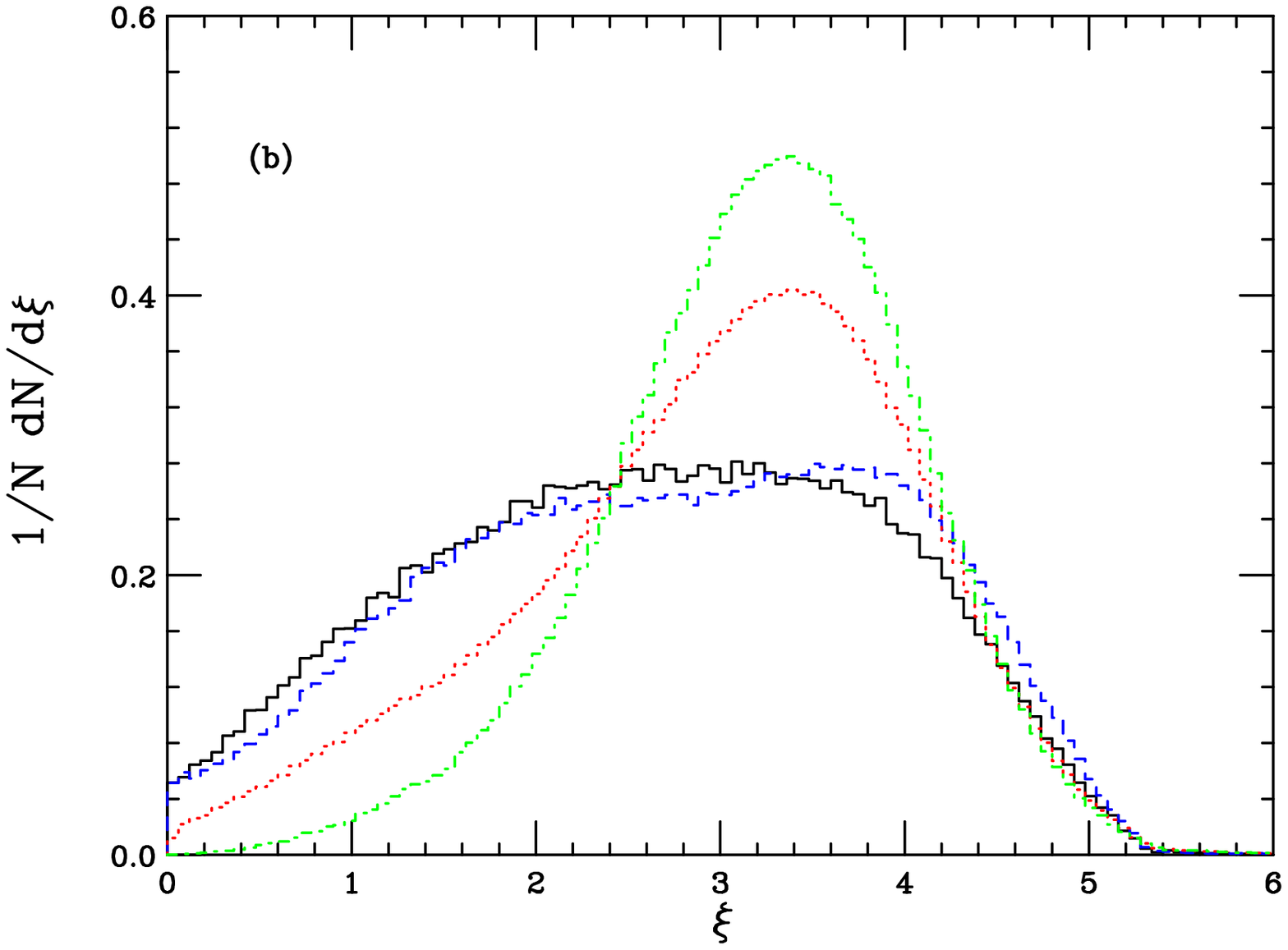}}}
\caption{$\xi$ spectrum for fixed and variable $L$. In the plots we 
follow the conventions adopted in Figs.~\ref{ptfix} and \ref{tetafix}.}
\label{xifix}
\end{figure}

A more evident impact of medium modifications is also exhibited 
by the $p_T$, $\theta$ and $\xi$ normalized spectra, presented in 
Figs.~\ref{ptfix}--\ref{xifix}. 
In fact, for $L=L_0$, 
the suppression at large $p_T$ and the enhancement at small $p_T$,
displayed by Fig.~\ref{ptfix}, is
stronger than what found when varying $L$ according to Eq.~(\ref{ell}).
The effect is milder for $\hat qL_0=2$~GeV$^2$, as we are starting from small
medium parameters, whereas it gets quite remarkable
for $\hat qL_0=50$~GeV$^2$ and $E=100$~GeV.
In the angular distributions (Fig.~\ref{tetafix}),
for $\hat qL_0=2$~GeV$^2$ the effect
of fixing $L$ is small but visible, while, for $\hat qL=50$~GeV$^2$, 
the fixed-length angular spectrum is above the variable-$L$ one for
$0.4<\theta<1.2$ and below at larger angles.
In fact, setting $L=L_0$, quarks and gluons always `see' 
a positive medium length, 
and partons, which in the variable-$L$ cascade would be potentially produced
at large $\theta$, are now allowed to further branch.
This explains why, for fixed $L$, we have a smaller {\it normalized} 
multiplicity
at very large angles and a higher one at middle $\theta$-values.
As all distributions are normalized to unity, the fixed-$L$ spectrum
is above the variable-$L$ one at small angles. 
Stronger suppression (enhancement) at small (large) values of 
$\xi$ is also exhibited by the energy-fraction plots in Fig.~\ref{xifix},
especially for $\hat qL_0=50$~GeV$^2$. 

\section{Conclusions}
We implemented medium-modified splitting functions in the HERWIG parton
shower algorithm, whose evolution satisfies the angular ordering
prescription.
Following the Q-PYTHIA implementation detailed in \cite{acs}, 
we added to the Altarelli--Parisi
splitting function a term depending on the medium properties, the
virtuality and the energy of the radiating parton. Such a modification was 
consistently implemented in both branching probability and Sudakov form 
factor. 

For the purpose of our phenomenological analysis, 
we included in HERWIG a fictitious process, where a shower is
originated by a single gluon with energy 10 and 100 GeV.
We studied the usual vacuum case, as well as showers in
a medium characterized by transport coefficient $\hat q$ and 
length $L$, with $L$ varying throughout the parton cascade.
As HERWIG showers satisfy angular ordering, when starting with just one
parton and not with a pair of colour-connected ones, we had to set by hand
the upper value of the evolution variable.

We ran HERWIG with modified splitting functions and, first of all, 
observed that the average parton multiplicity in a medium is higher 
than in the vacuum, with an enhancement which varies from 20\% to 80\%
when the accumulated transverse momentum $\hat qL_0$ runs between 2 and 50
GeV$^2$.
Then, we studied differential distributions, 
such as transverse momentum ($p_T$),
angle ($\theta$) and logarithmic energy-fraction ($\xi$) spectra.
Indeed, we found the features which one should expect in a 
dense medium with respect to the vacuum: enhancement at small $p_T$
and suppression at high $p_T$, more partons at large angles, 
suppression at low $\xi$, compensating an enhancement for
middle-high $\xi$ values. As for the results obtained for $E=10$~GeV,
we noticed a sharp peak at $p_T=\theta=\xi=0$, corresponding to
a fraction of events
with no emission at all. For this reason, we also
plotted the $\xi$ integrated distribution, which is insensitive to such a peak,
and found results more similar to the ones yielded the Q-PYTHIA code
and in qualitative agreement with the observations at RHIC.
We investigated the differential spectra for fixed values of the
length and found stronger medium effects, as in this case $L$ does
not decrease throughout the shower and is always positive.

As for the comparison with the results in \cite{acs}, overall we
found acceptable qualitative agreement, although, due to the fact that the
two shower algorithms are pretty different and that we have been considering
unphysical parton-level quantities, with hadronization switched off,
some quantitative discrepancies
are well visible. In fact, we observed that due to the choice of
the showering variables, more branchings are simulated in PYTHIA, 
at least when using the default parametrizations, even
before including medium modifications. In particular, sharp zero-peaks,
exhibited by all HERWIG distributions 
when considering a starting gluon of 10 GeV, were
not present in PYTHIA. 

We believe that it will be now very interesting to consider an actual 
physical process, e.g. at the LHC, investigate 
hadron-level observables and eventually
understand how, after including medium splitting functions, 
modified HERWIG fares with respect to Q-PYTHIA. 
For such a comparison to be trustworthy, both generators should be
tuned to the same data set.
Also, it will be very useful understanding whether the medium-induced
effects found in this analysis still persist once hadronization
is turned on.
As for the issue of
angular ordering and evolution variables, it will be cumbersome studying
quantities sensitive to colour coherence or non-global observables,
as in Refs.~\cite{cdf,bcd}, and understand whether the
discrepancies between HERWIG and PYTHIA still persist or 
medium-effect implementations wash them out.
However, as we said in the introduction, a comparison with
hadron-hadron and 
heavy-ion data is compulsory before making any strong statement on 
the role played by the shower evolution variable in a dense medium.
The documentation of the Q-HERWIG code, including medium-modified 
splitting functions, which should be not seen as an official release, but
rather as an add-up to the latest fortran version \cite{herwig},
is currently in progress \cite{qher}.

\section*{Acknowledgements}
We acknowledge M.H.~Seymour and 
U.A.~Wiedemann for discussions on these and related
topics. This work has
been supported by Ministerio de Ciencia e Innovaci\'on of Spain under
projects FPA2005-01963,
FPA2008-01177 and contracts Ram\'on y Cajal (N.A. and C.A.S.), 
by Xunta de Galicia
(Conseller\'ia de
Educaci\'on and Conseller\'ia de Innovaci\'on e Industria -- Programa Incite)
(N.A. and C.A.S.), by the
Spanish Consolider-Ingenio 2010 
Programme CPAN (CSD2007-00042) (N.A. and C.A.S.) and by the
European Commission grant PERG02-GA-2007-224770 and Xunta de Galicia
(Conseller\'\i a de Educaci\'on e Ordenaci\'on Universitaria) (C.A.S.).

\end{document}